\documentclass[pre,superscriptaddress]{revtex4}
\usepackage[utf8x]{inputenc}
\usepackage{ucs}
\usepackage{amsmath}
\usepackage{amsfonts}
\usepackage{amssymb}
\usepackage{makeidx}
\usepackage{graphicx}
\usepackage{xcolor}
\usepackage{listings}
\usepackage{subfig}

\begin{document}

\author{Jonas Andersen Bro} \affiliation{Roskilde University, Postbox
  260, DK-4000 Roskilde, Denmark}

\author{Kasper Sternberg Brogaard Jensen} \affiliation{Roskilde
  University, Postbox 260, DK-4000 Roskilde, Denmark}

\author{Alex Nygaard Larsen} \affiliation{Roskilde University, Postbox
  260, DK-4000 Roskilde, Denmark}

\author{Julia M. Yeomans} \affiliation{Oxford University, Rudolf Peierls
  Centre for Theoretical Physics, 1 Keble Road, Oxford, OX1 3NP, UK}

\author{Tina Hecksher} \affiliation{Centre Glass and Time, IMFUFA,
  Department of Science and Environment, Roskilde University, Postbox
  260, DK-4000 Roskilde, Denmark}

\title{The macroscopic pancake bounce}

\date{\today}

\begin{abstract}
  We demonstrate that the so-called {\em pancake bounce} of
  millimetric water droplets on surfaces patterned with hydrophobic
  posts [Nat. Phys. \textbf{10}, 515 (2014)] can be reproduced on
  larger scales. In our experiment, a bed of nails plays the role of
  the structured surface and a water balloon models the water
  droplet. The macroscopic version largely reproduces the features of
  the microscopic experiment, including the Weber number dependence
  and the reduced contact time for pancake bouncing. The scalability
  of the experiment confirms the mechanisms of pancake bouncing, and
  allows us to measure the force exerted on the surface during the
  bounce. The experiment is simple and inexpensive and is an example
  where front-line research is accessible to student projects.
\end{abstract}

\maketitle

\section{Introduction}
\begin{figure}
  \includegraphics[width=16cm]{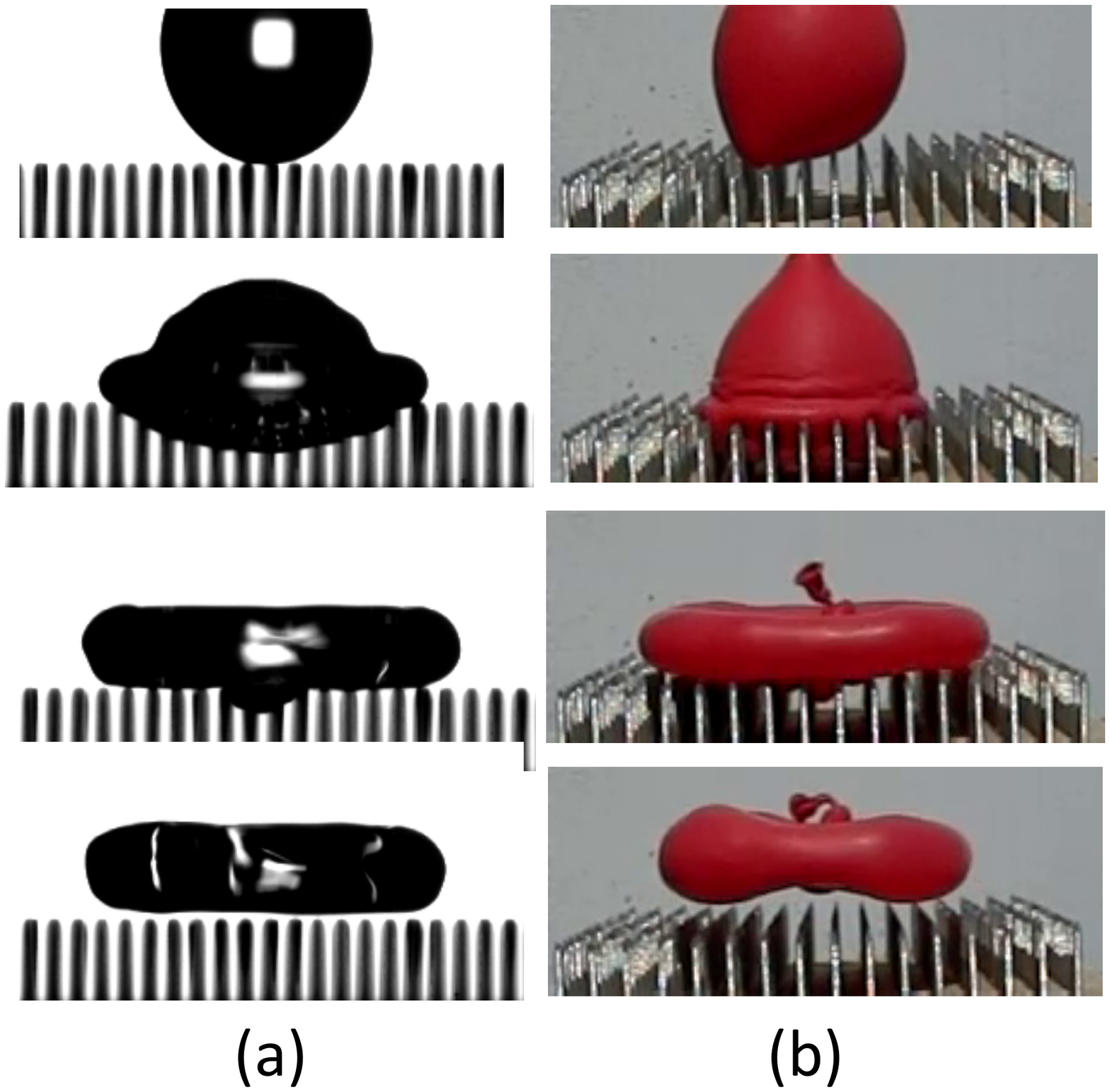}
  \caption{\label{microbounce} Rebound at maximum lateral extension
    (a) a millimetric droplet: the centre to centre spacing of the
    substrate posts is 200 $\mu$m, after \cite{Moevius2014}; (b) a balloon:
    the distance between nails is 1.85 cm}
\end{figure}
When a water drop hits a low friction, solid surface it typically
spreads and then retracts to its original radius before
bouncing. However Liu {\it et al.} \cite{Liu2014} have recently
demonstrated that an impinging drop of radius $\sim$ 1mm can leave a
substrate at its maximum extension, before retracting, and therefore
bounce in a pancake-like configuration (Fig.~\ref{microbounce}a). The
surfaces which result in so-called {\em pancake bouncing} are arrays
of hydrophobic posts of centre to centre spacing 200$\mu$m and height
800$\mu$m. Upon impact fluid is pushed between the posts, slowed, and
then expelled by the hydrophobic surfaces of the posts so that the
fluid entering and exiting the surface behaves like a spring. If the
fluid returns to the surface while the drop is at its maximum lateral
extension, and as long as it has sufficient energy, it is able to push
the drop off the surface in the pancake shape.

The present paper is based on a student project carried out at
Roskilde University \cite{Larsen2015} in the spring semester 2015. We
asked the question: can the pancake bounce be reproduced on a larger
scale? The idea was to model the water droplets with water-filled
balloons, where the rubber of the balloon mimics the surface tension
of the droplet, and to scale the structure of the surface
accordingly. The (surprising) answer was: yes, pancake bouncing is
observed for large balloons bouncing on a bed of nails at sufficiently
large impact velocities (Fig.~\ref{microbounce}b). A popular movie
\cite{pancakemovie} about the experiment is available.

We show that much of the microscopic phenomenology can be recreated at
macroscopic length scales. In particular we find that the contact time
of the balloon with the surface is independent of the impact velocity,
and reproduce a threshold in the impact velocity below which pancake
bouncing is suppressed. Moreover we are able to add to the microscopic
experiments by measuring the time-dependence of the force acting on
the surface as the balloon bounces.

In Section~\ref{experiments} we describe the experimental
details. Section~\ref{bouncing} compares how water balloons bounce on
a flat surface and the bed of nails, and discusses the forces exerted
on the substrates by the balloons as they bounce. We then discuss the
contact times in Section~\ref{contact}, and compare the results to
those obtained for millimetric water drops in
\cite{Liu2014}. Section~\ref{discussion} concludes the paper by
summarising our results and suggests directions for further work.

\section{Experimental details}
\label{experiments}
A digital reflex camera (Casio Exilim Pro EX-F1) capable of high-speed
recording up to 1200fps was used to produce movies of bouncing water
balloons. For the current purpose a frame rate of 300fps with
resolution 512x384 was sufficient to give enough details for
subsequent data analysis. Some movies were also shot at 600fps, but
the image resolution is lower (432x192). In addition, data on the
impact force were logged by an oscilloscope recording the voltage of a
piezo-electric disc placed under the bounce board.  Balloons were
ordinary ``party balloons'' purchased at the supermarket. Different
types were tested and the largest available were found to perform the
best.

Two different bounce boards were used: a flat board (flat surface) and
a nails board (spiked surface). The nails board was constructed to
give roughly the same relation between the radius of the balloon and
the distance between nails as in Ref.~\cite{Liu2014}. The nails (a
total of 256) are placed in a $1.85\text{ cm}\times1.85\text{ cm}$
square pattern.

\begin{figure}
  \includegraphics[width=8cm]{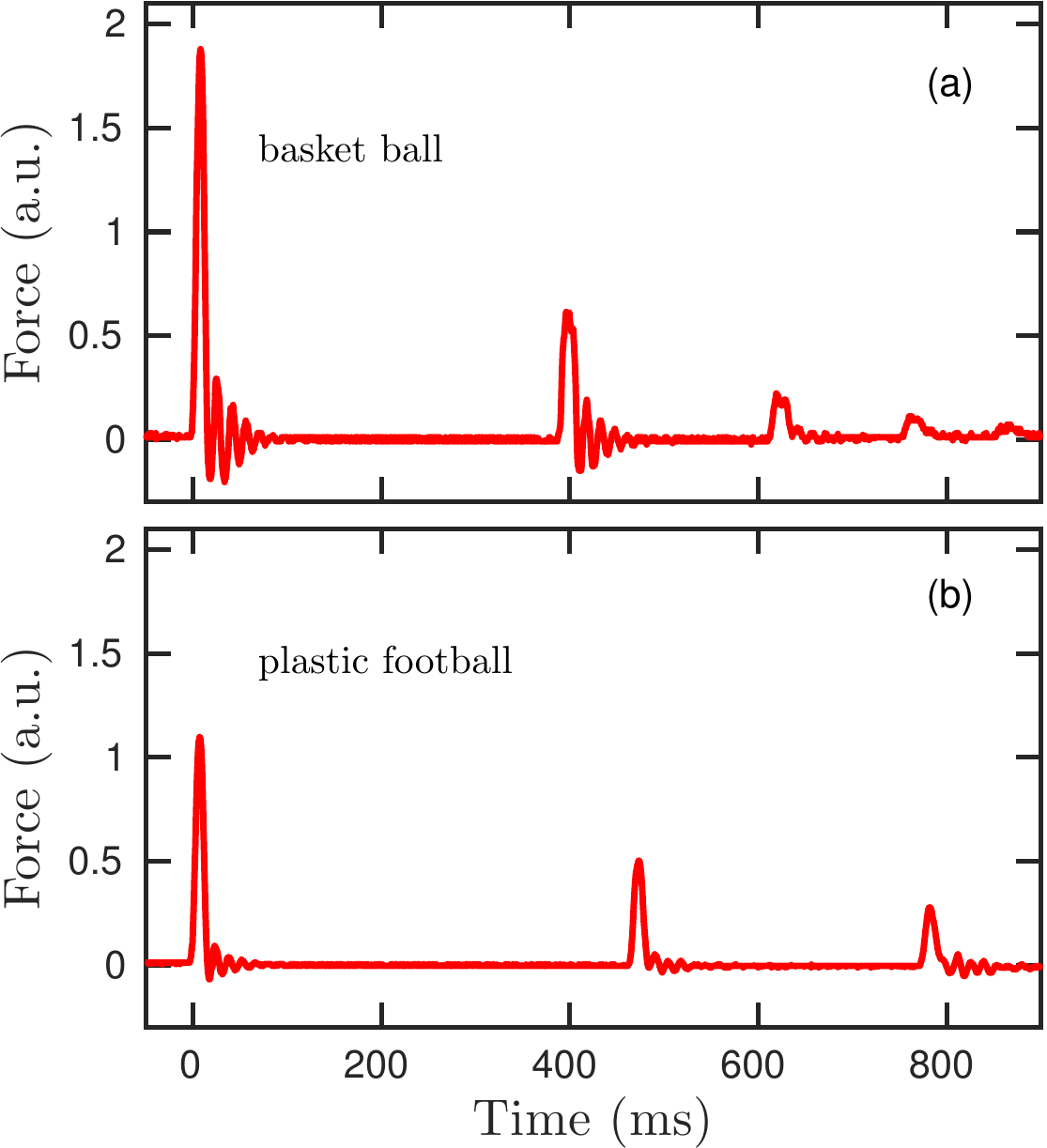}
  \caption{\label{scope_balls}Example of an oscilloscope data trace
    from the bounce of a basketball (a) and plastic football
    (b). Impact velocity in both cases was 3.4~m/s. The basket ball is
    heavier, so the impact force is relatively larger (peak is taller)
    than for the plastic ball. Also the subsequent ringing in the
    board (the oscillations seen after each impact) is more
    pronounced. The coefficient of restitution on the other hand is
    larger for the plastic ball, which is obvious from the time delay
    between the first and second bounce.}
\end{figure}

Figure \ref{scope_balls} shows an example of the data traces of the
oscilloscope for two ordinary air-filled balls (a basket ball and a
plastic football) rebounding from the flat bounce board. The two balls
have the same impact velocity. The basketball is however much heavier
which gives rise to a higher peak and to more ``ringing'' in the board
(oscillations following the bounce). It is also immediately seen that
the coefficient of restitution is smaller for the basketball since the
time differences between bounces are smaller than for the plastic
ball.

In order to be able to compare to droplet bouncing, we determined an
effective surface tension for the balloons by inflating them and
measuring the pressure, $\Delta P$.  The effective surface
tension $\gamma$ was then assumed to be defined by the Young-Laplace
equation \cite{Lautrup2005}
\begin{equation}
  \Delta P = \gamma\frac{2}{R}\,.
\end{equation}

The pressure was measured by a U-tube manometer and the balloon radius
$R$ from the circumference assuming spherical symmetry of the
balloon. We obtained $\gamma=60 \pm 30 \text{ Nm}^{-1}$ for both the
balloons at their impact radii. The large error bars are because the
effective surface tension varied between balloons and depended on
whether the measurement was made after the balloon was inflated or
deflated to the required radius \cite{Osborne1909,Hermans2010}.

\section{Results: Bouncing water balloons and substrate forces}
\label{bouncing}
\begin{figure*}
  \includegraphics[width=12cm]{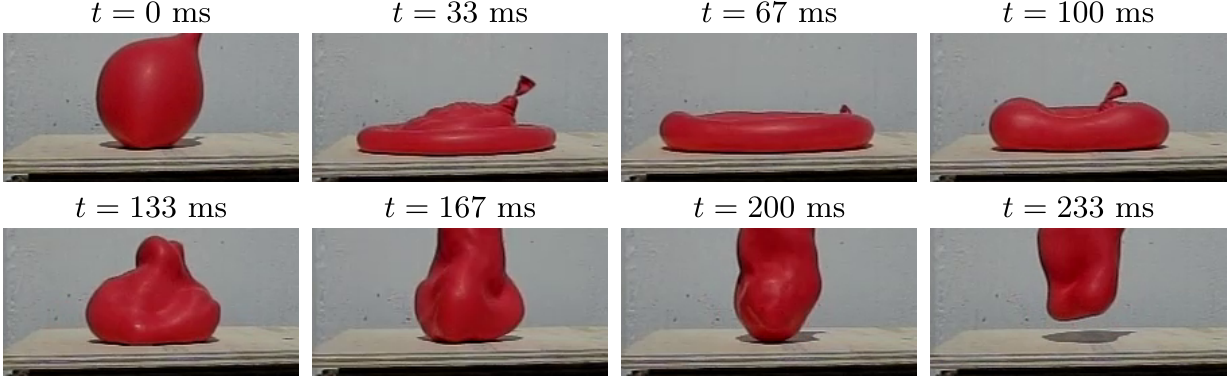}
  \caption{\label{normalbounce}Waterballoon bounce on a flat
    surface. The snapshots show the different stages of the
    bounce. The first contact with the surface defines the time
    $t=0$~ms. The balloon detaches from the surface at $t=210$~ms. The
    evolution and stages of the impact and bounce are nearly identical
    to those observed for millimetric water droplets, except that the
    bounce time for the water droplets $\sim$10~ms (for a comparison
    see e.g. \cite{Clanet2004}).}
\end{figure*}

Figure \ref{normalbounce} shows the different stages of an impact of
the 6~cm water balloon on a flat surface. Time $t=0$ is defined as the
first contact between balloon and surface. At around 66~ms the balloon
is maximally extended and starts retracting again, and between 200~ms
and 233~ms the balloon detaches from the surface. Except for the time
scale, the course of the bounce mimics closely what happens when a
water droplet of size $\sim$1~mm impacts on a hydrophobic surface
\cite{Clanet2004}.

\begin{figure*}
  \includegraphics[width=\textwidth]{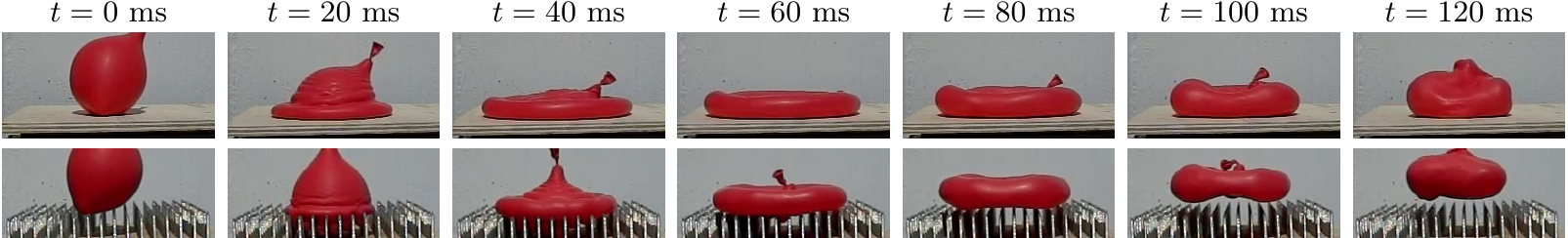}
  \caption{\label{pancakebounce}Comparison of a balloon bouncing on a
    flat surface and a spiked surface at the same impact velocity. The
    time evolution of the drop shape follows the same pattern in the
    two cases, except that the balloon detaches from the spiked
    surface at 65~ms (at the largest deformation) and then contracts
    in the air, while the balloon is in contact with the flat surface
    for much longer; it detaches at 210~ms after having contracted to
    an elongated cigar shape (compare Fig.\ \ref{normalbounce}).}
\end{figure*}

In Fig.\ \ref{pancakebounce} the bouncing of the water balloon on the
flat surface and on the bed of nails are compared at the same impact
velocity. The courses of the two impacts are initially
similar. However, in the latter case the balloon actually makes a
pancake bounce: it lifts off the bed of nails at its maximum
deformation and begins to retract in the air rather than on the
surface.
Figure~\ref{pancakebounce} also shows that the maximum extension is
smaller for the impact on the spiked surface.  This is because some of
the material penetrates into the nail pattern instead of being pushed
to the sides, and may also reflect an increased friction on the nails.

\begin{figure*}
\includegraphics[width=7cm]{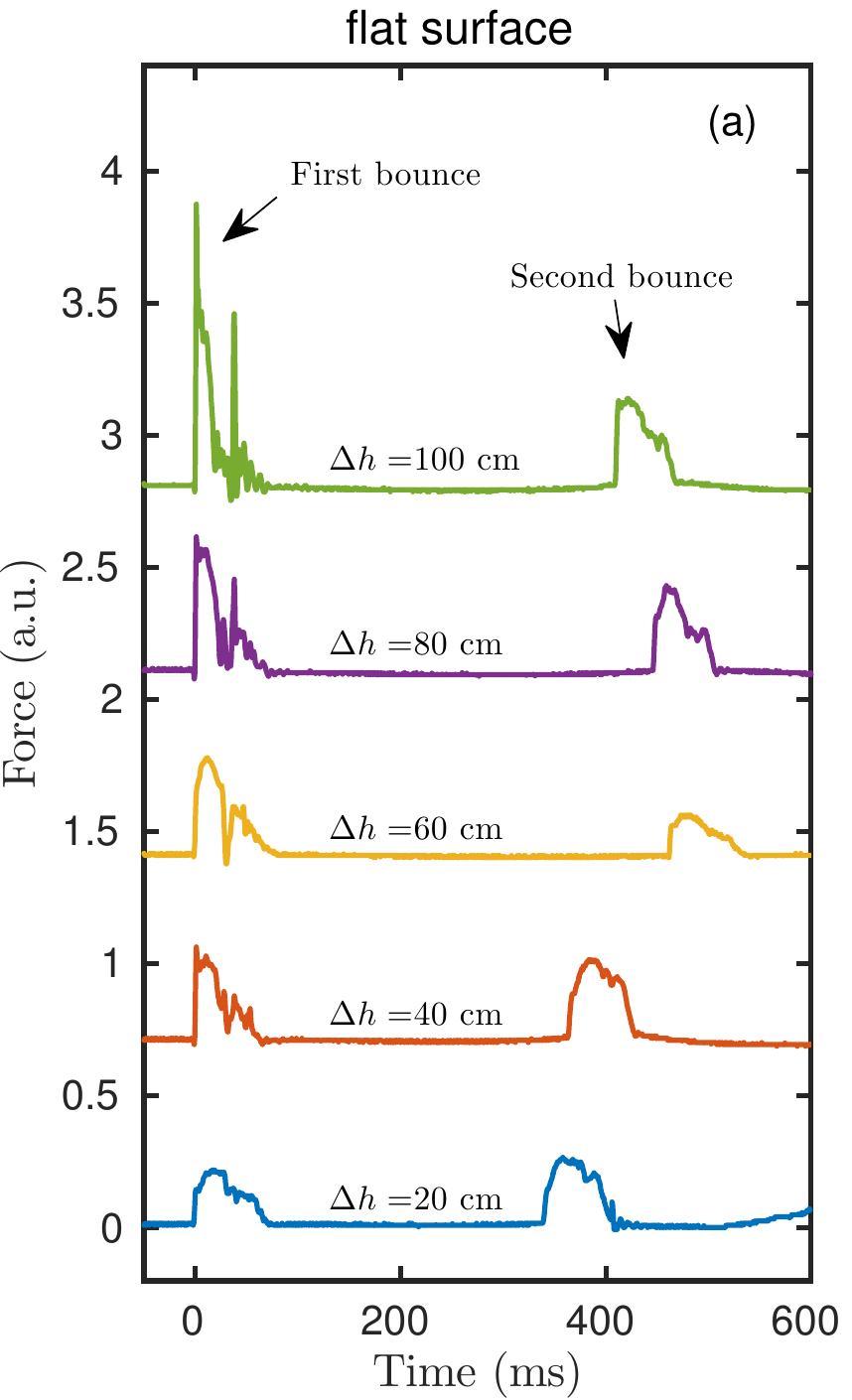}
\includegraphics[width=7cm]{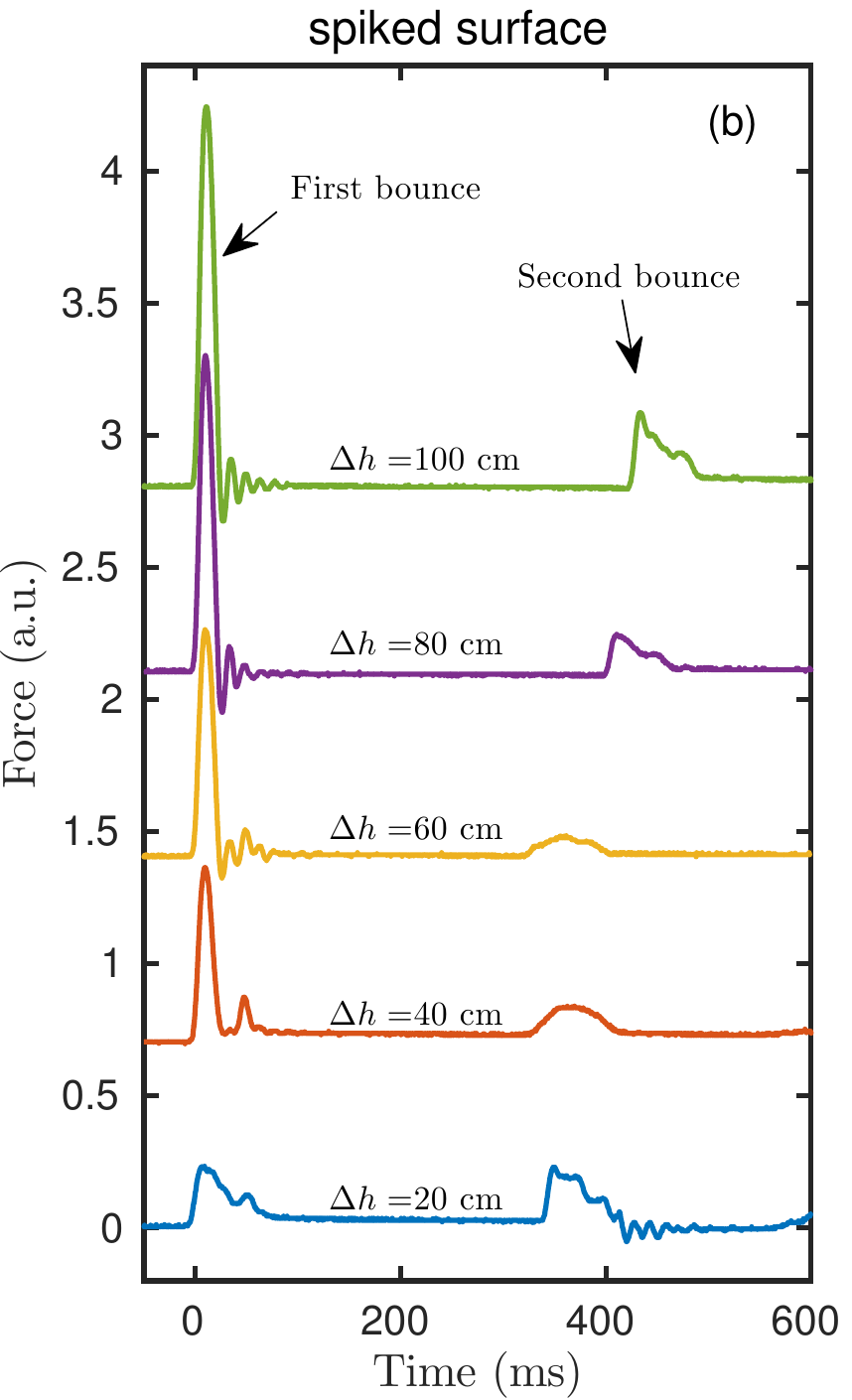}
\caption{\label{nb_series}Series of data traces of bounces with
  different impact velocities (shifted on the $y$-axis for
  clarity). The $\Delta h$ is the height from which the balloon was
  dropped (same $\Delta h$ implies same impact velocity). (a) Balloon
  bounces on a flat surface. The impact force has an asymmetric double
  peak (see zoom in Fig.\ \ref{compare}). (b) Balloon bounces on the
  spiked surface. The lowest two curves have a pronounced double-peak
  on first impact. From $\Delta h = 60$~cm the second peak
  disappears. This is the signature of pancake bouncing.}
\end{figure*}

In Fig.\ \ref{nb_series} oscilloscope traces (equivalent to force
curves) for a series of balloon bounces with different impact
velocities is shown for (a) a flat surface and (b) a spiked
surface. These results are for the 4.8 cm balloon. The normal force
during the impact on a flat surface has a characteristic asymmetric
double peak. There is a sharp increase as the balloon hits the
surface, then the force decreases as the balloon deforms and at
maximum deformation, the force is nearly zero. As the balloon starts
retracting, the force increases again; the balloon pushes off the
surface, and the center of mass is accelerated in the upwards
direction. Then, as the balloon leaves the balance, the force returns
to zero. This behaviour is shown on an expanded scale for a balloon
dropped from a height of 60~cm in Fig.~\ref{compare}a.

For the spiked surface, the low velocity impacts have a similar double
peak behaviour, but the shape is slightly different. This is because
the impact in this case is not as abrupt: some of the balloon and mass
penetrates into the nail pattern which softens the impact and gives a
force curve that is less steep initially. For high impact velocities,
however, there is a quantitative change: the first peak is sinusoidal
in shape, and the second peak disappears (see also
Fig.~\ref{compare}b). This behaviour of the force curves corresponds
to pancake bouncing: When the material that is forced into the nail
pattern recoils with sufficient energy, the balloon lifts off the
surface before it retracts.

\section{Results: contact times and comparison to water droplets}
\label{contact}
If any contribution due to dissipation can be neglected, the expansion
and contraction of the bouncing drop over the surface is controlled by
a balance between inertial forces, which act to spread the drop, and
surface tension, which acts to retract it. The dimensionless number
which controls the ratio of inertia and surface tension is the Weber
number We$=\rho v_0^2 R/\gamma$ where $\rho$ is the density of water
and $v_0$ is the impact velocity of the drop. 

The contact time
$t_\text{contact}$ is the time that the balloon (or droplet) is in
contact with the surface during the bounce. On dimensional grounds
\begin{equation}
  t_\text{contact} =c \sqrt{\rho R^3/\gamma}
\label{contacttime1}
\end{equation}
where $c$ is a numerical coefficient. Note that the contact time is
expected to be independent of the impact velocity. The physics behind
this is that the lateral motion during the bouncing approximates
simple harmonic motion, with a period independent of the velocity
amplitude. The scaling in Eq.~(\ref{contacttime1}) has been confirmed
for drops on a strongly hydrophobic surface \cite{Richard2002}.

\begin{figure}
\includegraphics[width=8cm]{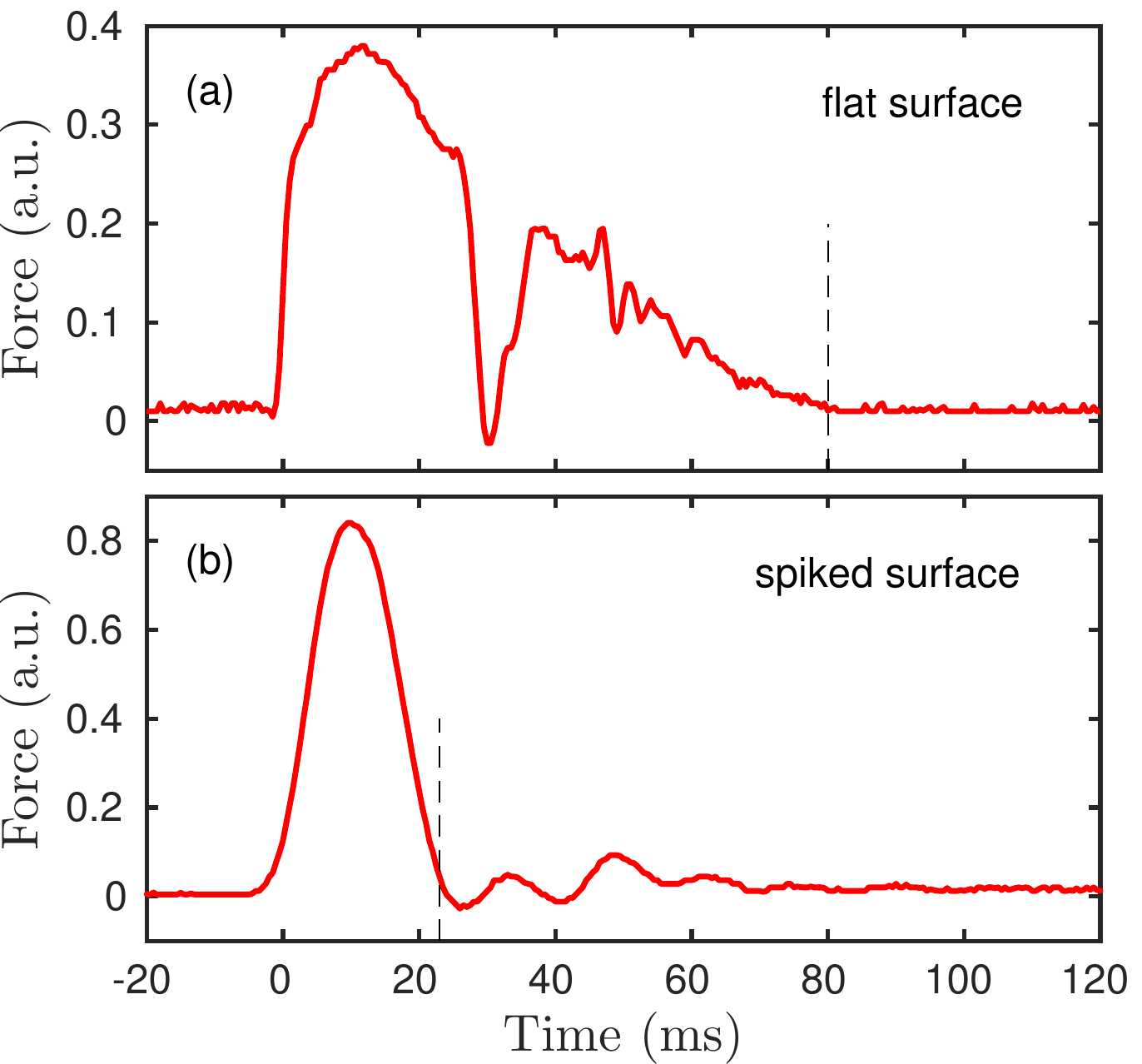}
\caption{\label{compare}Zoom on first impact of (a) a normal bounce
  and (b) a pancake bounce with the same impact velocity (same curves
  as Fig.\ \ref{nb_series} for $\Delta h = 60$~cm). Time when the
  balloon leaves the surface (equivalent to the contact time) is
  indicated by dashed vertical lines. Clearly, the contact time for a
  pancake bounce is reduced. But the shape of the peak is also
  markedly different: for the normal bounce there is a sharp increase
  in the force on impact and a double peak structure, whereas the
  pancake bounce corresponds to a single symmetric peak.}
\end{figure}
In our experiments the contact time of the bounce can be determined
from visual frame-by-frame inspection of the movies or from the scope
traces. The contact time in the latter case is taken as the width of
the impact peak: it starts at time zero when the scope registers the
onset of the impact and ends when the balloons detaches and the scope
again registers zero voltage.  The ringing of the board can make this
a little ambiguous, however if the oscillations are centered around
zero, we ascribe them to ringing; if not we assume that the balloon is
still in contact with the board. The procedure is illustrated in Fig.\
\ref{compare}. The same approach can be used to obtain contact times
for the second and third bounces if these are within the time window
and well resolved.

\begin{figure}
  \includegraphics[width=7.8cm]{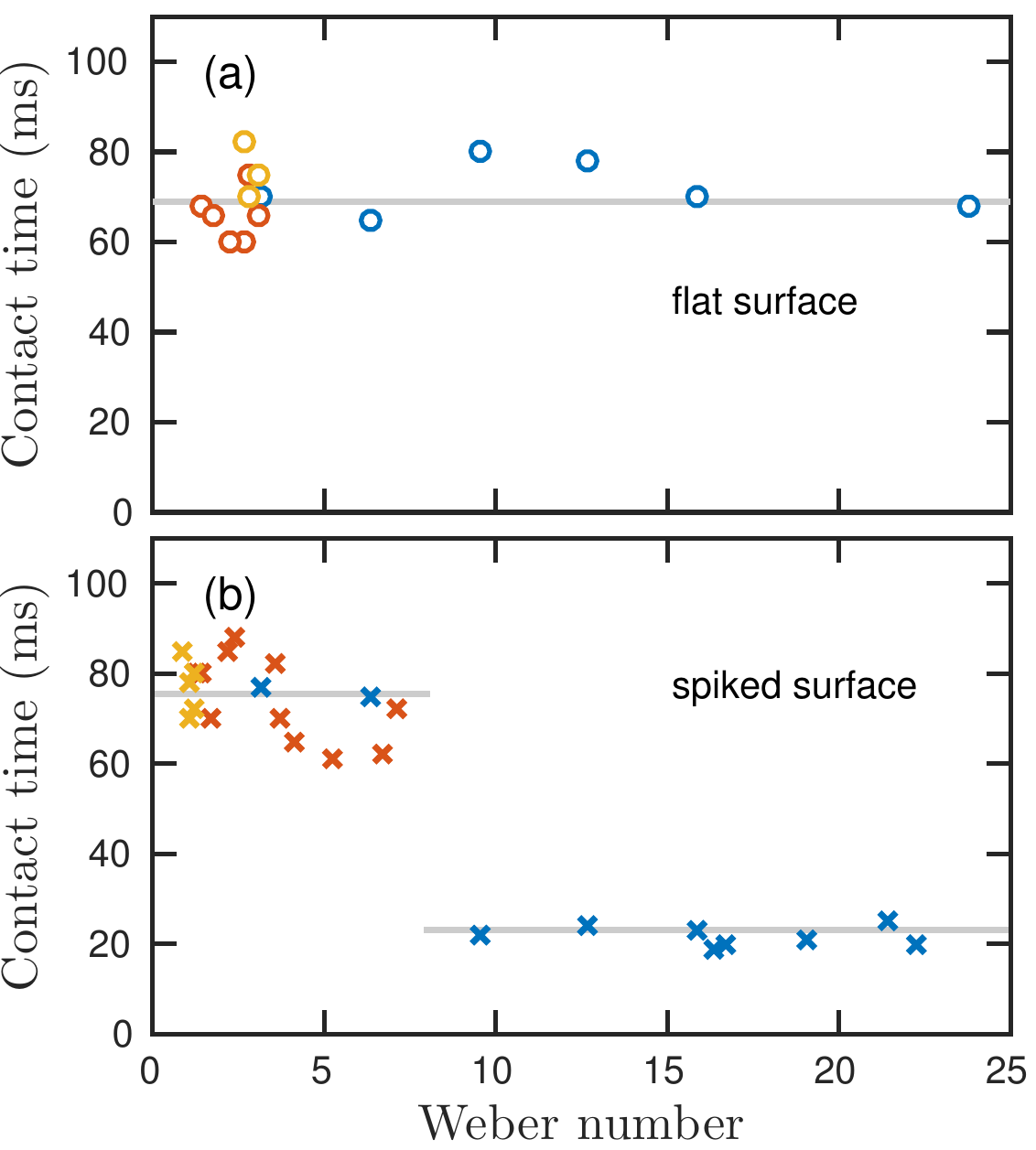}
  \includegraphics[width=8.5cm]{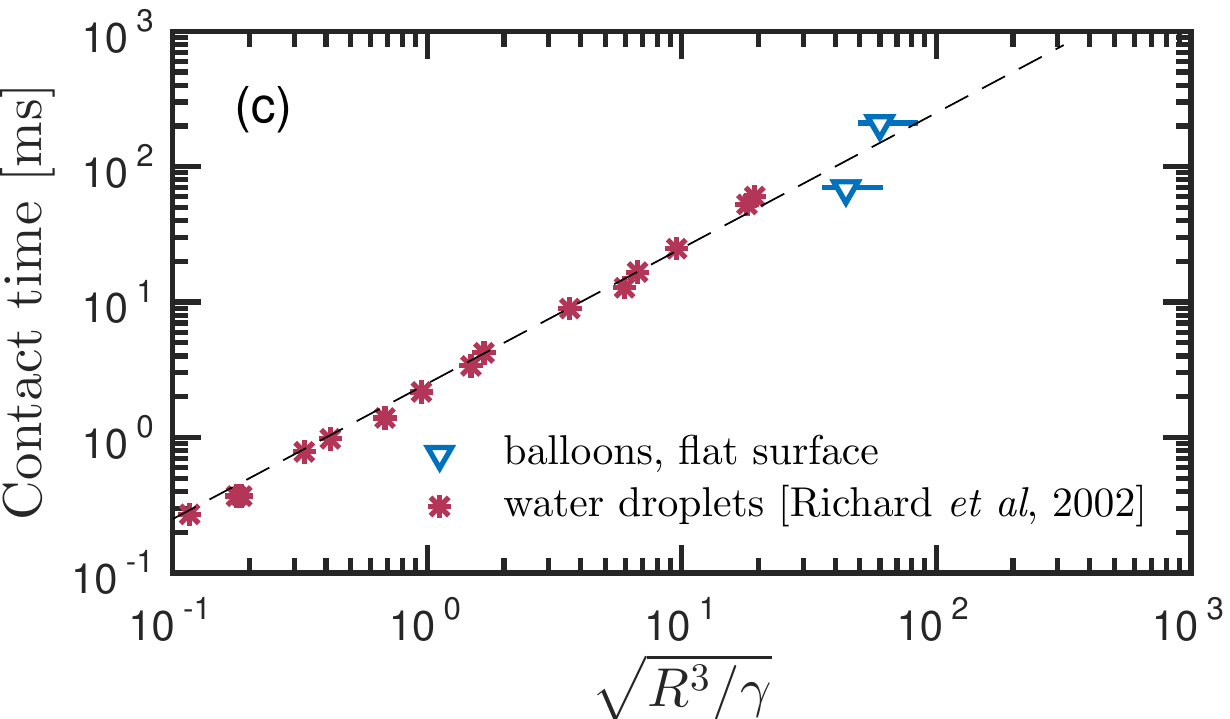}
  \caption{\label{contacttime} Contact time determined from the scope
    traces (see Fig.~\ref{compare}) as a function of Weber number for
    (a) a flat surface and (b) a spiked surface. Blue symbols are from
    first bounces, orange from second bounces, and yellow from third
    bounces (where available). Between $\text{We}=7$ and
    $\text{We}=9.5$ the contact time changes from around 80~ms to
    around 20~ms, a fourfold reduction. (c) Contact time on a flat
    surface compared to water droplets (data from
    \cite{Richard2002}). Error bars correspond to the estimated
    uncertainty of the balloon surface tension.}
\end{figure}

In Fig.\ \ref{contacttime} the contact time is shown as a function of
Weber number for bounces on (a) the flat surface and (b) the spiked
surface for the 4.8 cm balloon.  For the flat surface the data points
all lie around an average value of approximately 70~ms, independent of
impact velocity for a fixed balloon size, (i.e. independent of Weber
number) in agreement with results for water droplets
\cite{Richard2002}. For the spiked surface there is a change from a
constant value of 80~ms at low Weber numbers to a constant value of
around 20~ms at Weber numbers higher than $\sim 8$, marking the transition
from normal bouncing to pancake bouncing. This is consistent with the
behaviour observed by Liu \textit{et al} \cite{Liu2014} for a
microscopic surface.

To obtain a theoretical estimate of the contact time for pancake
bouncing we note that the force curve in Fig.~\ref{compare}b is
sinusoidal. Thus the fluid penetrating the substrate is behaving as a
harmonic spring. To estimate the force we assume that the balloon is
pinned on the nails and stretched into a spherical cap by the
downward-moving fluid. The resultant change in balloon surface area
for a cap of depth $z$ is $\pi z^2$. To obtain the total change in
area we multiply this by the number of pinning squares $\sim \pi
R^2/d^2$, where $d$ is the distance between nails, giving a stored
energy
$$
E=\gamma(\pi^2 R^2/d^2) z^2.
$$
Hence the force is
$$
F=-\gamma(2\pi^2 R^2/d^2) z= \rho (4 \pi R^3/3) \ddot{z}.
$$
Thus a half period, the time for filling and emptying the surface,
which for pancake bouncing is equivalent to the contact time, is
$$
t_\text{contact}
= 
\frac{d}{R} \sqrt{\frac{2\pi }{3}} \sqrt{\frac{ \rho R^3 }{\gamma}}
=0.56 \sqrt{\frac{ \rho R^3 }{\gamma}}.
$$
Using R=48 mm, $\gamma=60\text{ Nm}^{-1}$ gives a value for the contact
time of $24\text{ ms}$ in good agreement with the measured value, 20
ms. 

For this study we used only two different balloon sizes making it
difficult to test if the scaling relation~(\ref{contacttime1}) for the
contact time holds for the balloons. However, since the substance
inside the balloons is water, our results should be comparable to
water droplets. In Ref. \cite{Richard2002} such data were reported as
a function of drop radius. When comparing these data to the two
balloon data points we have used a value of $7.2\times 10^{-3}$~N/m
for the surface tension of water. As shown in Fig.\
\ref{contacttime}(c), the balloon data lie nicely in continuation of
the results for water droplets within the
uncertainty.  

\begin{figure}
  \includegraphics[height=4.7cm]{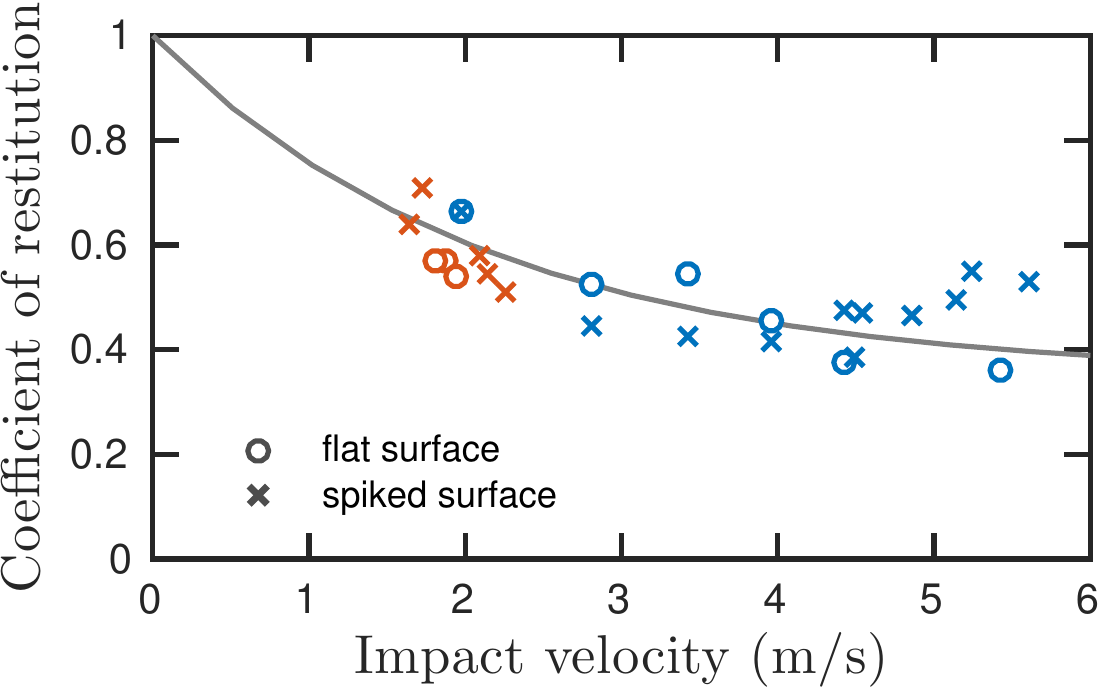}
  \caption{\label{COR} Coefficient of restitution defined as ratio
    between velocity before and after impact as a function of impact
    velocity. First bounces are in blue, second bounces in orange. The
    line is a guide to the eye.}
\end{figure}

Lastly, we looked at the coefficient of restitution of the balloon
bounces. The coefficient of restitution (COR) is defined as the ratio
between the speed immediately after and the speed immediately before
the impact and is a measure of the energy loss during impact. If the
impact is perfectly elastic COR is identically one, while for a
perfectly inelastic impact COR is zero. As a function of impact
velocity this is usually a curve that is close to one at low impact
velocities then decreases to level off at a constant value at high
impact velocities. In Fig.\ \ref{COR} we plot COR as a function of
impact velocity which follows the expected pattern. Results from
bounces on both flat surface (circles) and spiked surface (crosses)
are shown and it seems that there is no significant difference between
the two types of bounces.

\section{Concluding remarks}
\label{discussion}

We have studied water-filled balloons impacting on a flat surface and
on a bed of nails. On flat surfaces the balloons spread, retract and
then bounce with a contact time independent of the impact velocity. On
the nail surface the behaviour is similar at low Weber
numbers. However at high Weber numbers the balloon leaves the nails
close to its maximum extension, in a pancake shape. The contact time
for pancake bouncing is reduced over that for a flat surface by a
factor $\sim 4$.

Force balance measurements indicate a double peaked structure for a
normal bounce, with maxima associated with the the drop hitting and
leaving the surface. For a pancake bounce there is a single, symmetric
peak of a sinusoidal form. We argue that the harmonic force results
from the balloon being pushed down between the posts by the impacting
fluid, and then acting as a spring to launch the drop before
retraction.

The behaviour of the balloons is surprisingly similar to that of
millimetric bouncing drops, but with timescales longer by a factor
$\sim$ 10. In particular pancake bouncing has been observed for
substrates patterned with hydrophobic posts with a similar reduction
in the contact time. However, here the spring force is provided by the
hydrophobic covering of the posts which tends to decelerate and then
eject fluid entering the surface.

The experiment is accessible to undergraduate students in terms of
expertise, cost and understanding. In the future it might be of interest
to probe the analogy between water droplets and water filled balloons
in more detail by considering a greater range of balloon dimensions or
higher Weber numbers, when drops break up upon bouncing but balloons
cannot.



\end{document}